\documentclass[twocolumn]{aastex61}
\usepackage{graphicx}				
\usepackage{xcolor}
\usepackage[sort&compress]{natbib}
\usepackage[hang,flushmargin]{footmisc}
\usepackage[counterclockwise]{rotating}
\usepackage{amsmath,natbib,graphicx}
\usepackage{lipsum}
\usepackage{gensymb}

\shortauthors{Hayworth et al.}
\usepackage{wrapfig}
\shorttitle{Warming Early Mars with Climate Cycling}
\usepackage{enumitem,xcolor}

\begin{document}
\graphicspath{ {./} }
\DeclareGraphicsExtensions{.pdf,.eps,.png}

\title{WARMING EARLY MARS WITH CLIMATE CYCLING: THE EFFECT OF $CO_{2}$-$H_{2}$ COLLISION-INDUCED ABSORPTION}
\published{by Icarus July 15 2020}

\author{Benjamin P.C. Hayworth}
\affiliation{Department of Geosciences, Pennsylvania State University}

\author{Ravi Kumar Kopparapu}
\affiliation{NASA Goddard Space Flight Center}

\author{Jacob Haqq-Misra}
\affiliation{Blue Marble Space Institute of Science}

\author{Natasha E. Batalha}
\affiliation{University of California, Santa Cruz}

\author{Rebecca C. Payne}
\affiliation{Department of Geosciences, Pennsylvania State University}

\author{Bradford J. Foley}
\affiliation{Department of Geosciences, Pennsylvania State University}

\author{Mma Ikwut-Ukwa}
\affiliation{Department of Astronomy, Harvard University}

\author{James F. Kasting}
\affiliation{Department of Geosciences, Pennsylvania State University}
  
\correspondingauthor{Benjamin Hayworth}
\email{bph8@psu.edu}


\begin{abstract}
Explaining the evidence for surface liquid water on early Mars has been a challenge for climate modelers, as the sun was $\sim$30\% less luminous during the late-Noachian. We propose that the additional greenhouse forcing of $CO_{2}$-$H_{2}$ collision-induced absorption is capable of bringing the surface temperature above freezing and can put early Mars into a limit-cycling regime. Limit cycles occur when insolation is low and $CO_{2}$ outgassing rates are unable to balance with the rapid drawdown of $CO_{2}$ during warm weathering periods. Planets in this regime will alternate between global glaciation and transient warm climate phases.  This mechanism is capable of explaining the geomorphological evidence for transient warm periods in the martian record.  Previous work has shown that collision-induced absorption of $CO_{2}$-$H_{2}$ was capable of deglaciating early Mars, but only with high $H_{2}$ outgassing rates (greater than $\sim$600 Tmol/yr) and at high surface pressures (between 3 to 4 bars). We used new theoretically derived collision-induced absorption coefficients for $CO_{2}$-$H_{2}$ to reevaluate the climate limit cycling hypothesis for early Mars. Using the new and stronger absorption coefficients in our 1-dimensional radiative convective model as well as our energy balance model, we find that limit cycling can occur with an $H_{2}$ outgassing rate as low as $\sim$300 Tmol/yr at surface pressures below 3 bars. Our results agree more closely with paleoparameters for early martian surface pressure and hydrogen abundance. 
\end{abstract}

\section{Introduction}

	Understanding how Mars' early atmosphere compensated for the Sun's decreased luminosity is fundamental to explaining surface features that formed during the earliest part of the planet's history. Ample evidence exists for substantial liquid water on Mars' surface during the late Noachian and early Hesperian. This includes river channels \citep{doi:10.1029/2009JE003548}, long-lived standing lakes \citep{Grotzingeraac7575}, and phyllosilicates formed by aqueous alteration of other minerals \citep{poulet2005phyllosilicates}. Ground-based observations of the ratio of deuterium to hydrogen (D/H) in martian ground ice suggest a global equivalent water layer at least 137 m deep \citep{villanueva2015strong}. Mars' actual initial surface water content could have been larger than this if much of the H was lost through hydrodynamic escape, removing D along with H and increasing the fractionation factor \citep{batalha2015testing}. The existence of sustained surface water requires that early Mars' surface temperatures must have been locally and/or episodically above 273 K. However, the Sun was approximately 75\% as bright at 3.8 Ga \citep{gough1981solar,bahcall2001solar,feulner2012faint}, and so the solar flux at 1.52 AU would have been only $\sim32\%$ that of modern Earth. As a result, it is impossible to keep Mars' average surface temperature above 273 K using just $CO_{2}$ and $H_{2}O$ as greenhouse gases \citep{kasting1991co2,wordsworth2013global}.
	
	The difficulties in modeling a sustained warm climate on early Mars have led some to suggest that the surface of Mars has always been frozen, except in the immediate aftermath of giant impact events \citep{segura2002environmental,segura2008modeling,segura2012impact}. But such 'cold early Mars' hypotheses have difficulties explaining the formation of the ancient river valleys seen on Mars today \citep{hoke2011formation,steakley2019testing,turbet2020environmental}. Episodic warming from punctuated volcanic outgassing of $SO_{2}$ \citep{halevy2014episodic} also has problems working quantitatively because the oxidation of $SO_{2}$ forms reflective sulfate aerosols which cool the climate \citep{tian2010photochemical,kerber2015sulfur}.
	
	These difficulties can be overcome if the early martian atmosphere was rich in $H_{2}$. $H_{2}$ lacks a permanent electric dipole moment, so it is not a conventional greenhouse gas, but it can gain an induced dipole moment through collisions with other molecules \citep{birnbaum1978far,birnbaum1996collision}. Quantitative studies have suggested that early Earth may have been warmed in part through collision-induced absorption (CIA) from interactions between $H_{2}$ and $N_{2}$  \citep{wordsworth2013hydrogen}. This absorption extends through the 8-12 $\mu$m window region, enabling $H_{2}$ to act as an effective greenhouse gas relative to $H_{2}O$ and $CO_{2}$. 

	\citet{ramirez2014warming} assumed that collisional excitation of $H_{2}$ by $CO_{2}$ would be as efficient as excitation by $N_{2}$ and applied the $N_{2}$-$H_{2}$ CIA coefficients to early Mars-like conditions using a 1-D climate model. They found that 20\% $H_{2}$ in a 1.3-bar $CO_{2}$ atmosphere, or 5\% $H_{2}$ in a 3-bar $CO_{2}$ atmosphere, was required to keep global mean average surface temperature above 273 K. This $H_{2}$ could have been supplied by direct volcanic outgassing from Mars' highly reduced mantle -  $fO_{2}$ 3 log units lower than the terrestrial value \citep{wadhwa2001redox}, as well as from serpentinization, photochemical oxidation of other reduced volcanic gases \citep{batalha2015testing}, large impacts \citep{haberle2019impact}, and anaerobic iron oxidation \citep{tosca2018magnetite}. Carbon may have been outgassed initially as $CO$ or $CH_{4}$, but it would have been photochemically oxidized to $CO_{2}$ within a relatively short time frame  \citep{batalha2015testing}.

	The rates at which $CO_{2}$ and $H_{2}$ were supplied to Mars' early atmosphere, as well as the rate at which hydrogen escaped to space, are both poorly known \citep{wordsworth2017transient}. Both supply rates presumably declined with time as Mars' interior cooled and volcanic activity subsided. This may have put Mars into a 'limit cycling' climate regime in which surface temperatures oscillated between warm and globally glaciated conditions. Such behavior is predicted when stellar insolation is low \citep{menou2015climate} and/or when $CO_{2}$ outgassing rates are slow \citep{tajika2003faint,kadoya2019outer,abbot2016analytical}. Both conditions would likely have been satisfied during parts of Mars' early history. During periods of global glaciation, when silicate weathering is slow or absent entirely, outgassing causes atmospheric $CO_{2}$ to accumulate, increasing surface temperatures until deglaciation can occur. $H_{2}$ increases along with $CO_{2}$, providing the additional warming needed to make this happen \citep{batalha2016climate}. During interglacial periods, the rate of $CO_{2}$ drawdown from weathering outpaces the rate of $CO_{2}$ outgassing, and so the cycle repeats. This cycling depends on there being enough weatherable material  for the weathering to not be supply-limited (See Discussion~\ref{sec:discussion2}). Our own energy-balance climate model (EBM) predicts long, cold periods interspersed by transient warm periods lasting 10 Myr each (ibid.). 
	
	Recently, \citet{wordsworth2017transient} published new, theoretically calculated CIA coefficients for $CO_{2}$-$H_{2}$ and $CO_{2}$-$CH_{4}$ collisions and tested them in 1-D simulations of early Mars climate conditions. They showed that $CO_{2}$-$H_{2}$ and $CO_{2}$-$CH_{4}$ CIA both fill a gap in the 'atmospheric window' region of $CO_{2}$-rich atmospheres, between 250 and 500 cm-1 (20-40 $\mu$m), where absorption of IR radiation is weak. (The normal 8-12 $\mu$m 'window' that exists in Earth's atmosphere is not present in dense $CO_{2}$ atmospheres because of the absorption caused by the 9.4- and 10.4-$\mu$m 'hot' bands of $CO_{2}$.) The coefficients for $CO_{2}$-$H_{2}$ CIA were calculated to be stronger than those for $N_{2}$-$H_{2}$ CIA because the more heterogeneous electron density distribution of $CO_{2}$ compared to $N_{2}$ creates stronger multipole moments and polarizability. 
	
	Even more recently, an experimental study by \citet{turbet2019far} indicated that $CO_{2}$-$H_{2}$ CIA is weaker in the 300-550 cm-1 band than calculated by \citet{wordsworth2017transient}. As both the CIA coefficients from \citet{turbet2019far} will result in weaker CIA - they should increase the overall $H_{2}$ needed to provide warming. We will include their coefficients in our future work. This means that the results we calculate using \citet{wordsworth2017transient} is an upper-limit on the amount of $CO_{2}$-$H_{2}$ warming we can expect, with the CIA coefficients from \citet{turbet2019measurements} being the other end-member. Our goal is to determine whether using these coefficients will allow us to significantly lower the outgassing rates needed to sustain limit cycling behavior.

\section{Methods}
\label{sec:methods}

	We used a 1-D radiative-convective model, CLIMA \citep{ramirez2014warming,batalha2016climate} and an EBM, named HEx \citep{haqq2016limit,batalha2016climate}, to study this problem. The first model, CLIMA, is computationally inexpensive and can cover large regions of parameter space. We use it to find qualitative trends for different outgassing rates and to help us estimate where limit cycling is likely to occur. It is also able to find the lower limit on $H_{2}$ needed to deglaciate the planet. We then use the second model, HEx, to quantitatively test this region of parameter space and compare to the predictions made by CLIMA. These models are described briefly below. More details can be found in the original papers.

\subsection{CLIMA (1-D radiative-convective model)}
\label{sec:methods1}

	The radiative transfer parameterizations of the CLIMA model used by \citet{batalha2016climate} were updated to include the new set of $CO_{2}$-$H_{2}$ CIA coefficients \citep{wordsworth2017transient}. Using CLIMA, we calculated a total of 27,300 outgoing longwave radiation (OLR) fluxes for differing values of surface pressure, surface temperature, and mixing ratio of $H_{2}$. For each specified condition, the code calculates profiles of atmospheric pressure, temperature, and $H_{2}$ mixing ratio. All gas other than $H_{2}$ is assumed to be $CO_{2}$ or $H_{2}O$. A fully saturated moist adiabat is used to estimate $H_{2}O$ concentrations in the troposphere. The $H_{2}O$ mixing ratio is held constant above the tropopause. Using these profiles, the code calculates radiative fluxes in each of 55 wavelength intervals. These are summed to compute the total outgoing long-wavelength flux at the top of the atmosphere \citep{kasting1988runaway}. Similarly, we calculated a total of 546,000 planetary albedo (PALB) values for differing surface pressures, surface temperatures, $H_{2}$ mixing ratios, zenith angles, and surface albedos. The OLR and PALB values were calculated using polynomial fits over the previously listed physical variables, similar to the procedure used by \citet{batalha2016climate}. A more detailed description of the fitting procedure, and expected error of the fits, can be found in the Appendix.
	
\subsection{HEx (Energy-balance model)}
\label{sec:methods2}

	We used an energy-balance model, HEx (the Habitability EBM for Exoplanets), to explore climate limit cycling behavior. Such models are intermediate in complexity between 1-D radiative-convective models (such as CLIMA) and 3-D GCMs \citep{wordsworth2013global,soto2015martian}. Unlike a 1-D model, an EBM can include latitudinal heat transport, seasonal changes, fractional land/ocean cover, and ice-albedo feedback. Additionally, because most of the physics is heavily parameterized, and because it operates in one dimension (latitude), an EBM can be used to explore climate evolution on geologic timescales. This is difficult to accomplish using a 3-D GCM because the required run times would be too long (though not impossible, as previous authors have explored these regimes by coupling 0-D and 3-D model components to reach the required timescales \citep{paradise2017gcm}). The radiative flux and albedo parameterizations of the EBM used by \citet{batalha2016climate} and \citet{haqq2016limit} were updated to explore the effect of the new $CO_{2}$-$H_{2}$ CIA coefficients on limit cycling behavior. This model also includes geochemical cycling of greenhouse gases as well as a parameterization of ice-albedo feedback. 
	
	HEx calculates the average temperature as a function of latitude according to:
\\
\begin{equation}
C\frac{\delta T}{\delta t} = \overline{S(\theta)}(1 - \alpha_{p}) - F_{OLR} + \frac{1}{cos(\theta)}\frac{\delta}{\delta \theta}(Dcos(\theta)\frac{\delta T}{\delta \theta})
\end{equation}
\\
Here, $\overline{S(\theta)}$ is the stellar flux as a function of latitude, relative solar constant, obliquity, and orbital eccentricity. Calculated using CLIMA, $\alpha_{p}$ is the planetary albedo and $F_{OLR}$ is the flux of outgoing longwave radiation (See Section~\ref{sec:methods1}). $C$ is the effective heat capacity of the surface and atmosphere, and $D$ is our meridional heat diffusion constant which is parameterized as a function of atmospheric composition, surface pressure, and rotational rate of the planet, following \citet{williams1997habitable}. We used 18 latitude bands in our model. 

	The carbon cycle is modeled following \citet{menou2015climate}:
\\
\begin{equation}
\frac{W}{W_{\bigoplus}} = \underbrace{( \frac{pCO_{2}}{p_{\bigoplus}})^{\beta}}_{A} \times \overbrace{e^{k_{act}(T_{s} - 288)}(1 + k_{run}[T_{s} - 288])^{0.65}}^{B}
\end{equation}
\\
Here, $W_{\bigoplus}$ represents the weathering rate on the modern Earth, which has a mean temperature of 288 K. For modern Earth, this must equal the $CO_{2}$ volcanic outgassing rate, $\sim7.5$ Tmol/yr \citep{jarrard2003subduction}. Here, it is treated as a free parameter. Term $A$ represents the dependence of the weathering rate on $pCO_{2}$. We assume $\beta$ = 0.5, although it can theoretically range from 0 - 1.0 \citep{berner1992weathering}. We tuned the model to the modern biotic Earth, $p_{\bigoplus}$ = 1$\times$$10^{-2}$ bar, to remain consistent with the results of \citet{batalha2016climate} and \citet{haqq2016limit}. This value is significantly higher than the preindustrial atmospheric $pCO_{2}$ ($\sim$3$\times$$10^{-4}$ bar) because of enhancement of soil $pCO_{2}$ by vascular plants. This is a large point of uncertainty in our parameterization of weathering, and a more detailed discussion is given in \citet{batalha2016climate}. The uncertainty in weathering rates compliments our uncertainty in outgassing rates, which is why it is treated as a free parameter. Term $B$ represents the temperature dependence of the weathering rate, which has an assumed activation energy, kact = 0.09 $K^{-1}$, and a runoff efficiency factor, $k_{run}$ = 0.045 $K^{-1}$ \citep{berner1992weathering}. 

	We assume the hydrogen budget is controlled by the volcanic flux of hydrogen and by diffusion-limited escape to space \citep{walker1981negative} (though it potentially could have escaped slower than the diffusion-limit - See Section~\ref{sec:discussion4}):  
\\
\begin{equation}
\phi_{esc}(H_{2}) = \frac{b_{i}}{H_{a}} \frac{fH_{2}}{1 + fH_{2}}
\end{equation}
\
  \begin{figure}[!]
\begin{center}
 \includegraphics[width=3.6in]{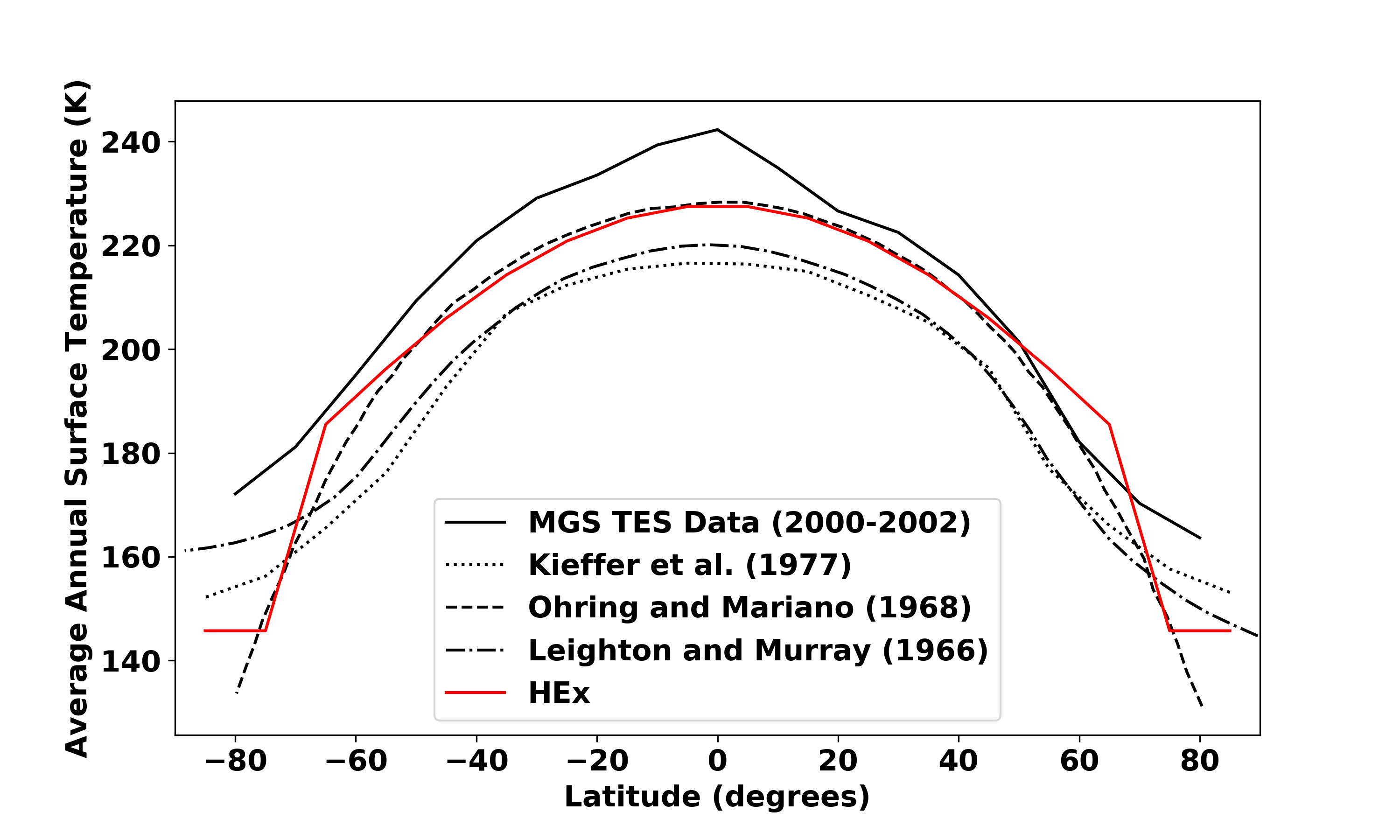}
 \end{center}
 \caption{The average annual surface temperature as a function of latitude for modern Mars. The dash-dot curve corresponds to the distribution from \citet{leighton1966behavior}. The dashed curve corresponds to the distribution from \citet{ohring1968seasonal}. The dotted curve corresponds to the distribution from \citet{kieffer1977thermal}. The solid line was taken from us latitudinally averaging MGS TES data from 2000-2002 over one Martian year. The red curve is from HEx for modern Martian conditions.}
 \label{fig:compare}
\end{figure}
We treated the escape rate in the same way as \citet{batalha2016climate}, setting $\frac{b_{i}}{H_{a}} = 1.6 \times 10^{13} cm^{-2}s^{-1}$, a value appropriate for $H_{2}$ diffusing through $CO_{2}$. Because the escape process is relatively fast, the mixing ratio of $H_{2}$ remains directly proportional to the outgassing rate, regardless of surface pressure. The $H_{2}$ and $CO_{2}$ outgassing rates were treated as a free parameters.
	As a benchmark for HEx, we used modern martian input values ($pCO_{2}$, orbital parameters, solar constant, surface albedo, etc.) to see how our latitudinal dependence on average annual surface temperature compared to some distributions found in the literature. See Figure 1. The vertical spread between sources is due to modeling differences (i.e. assumptions of surface pressure, atmospheric composition, thermal inertia, albedo, etc.).

\section{Results}
\label{sec:results}

\subsection{CLIMA}
\label{sec:results1}

  \begin{figure}[h!]
\begin{center}
 \includegraphics[width=3.5in]{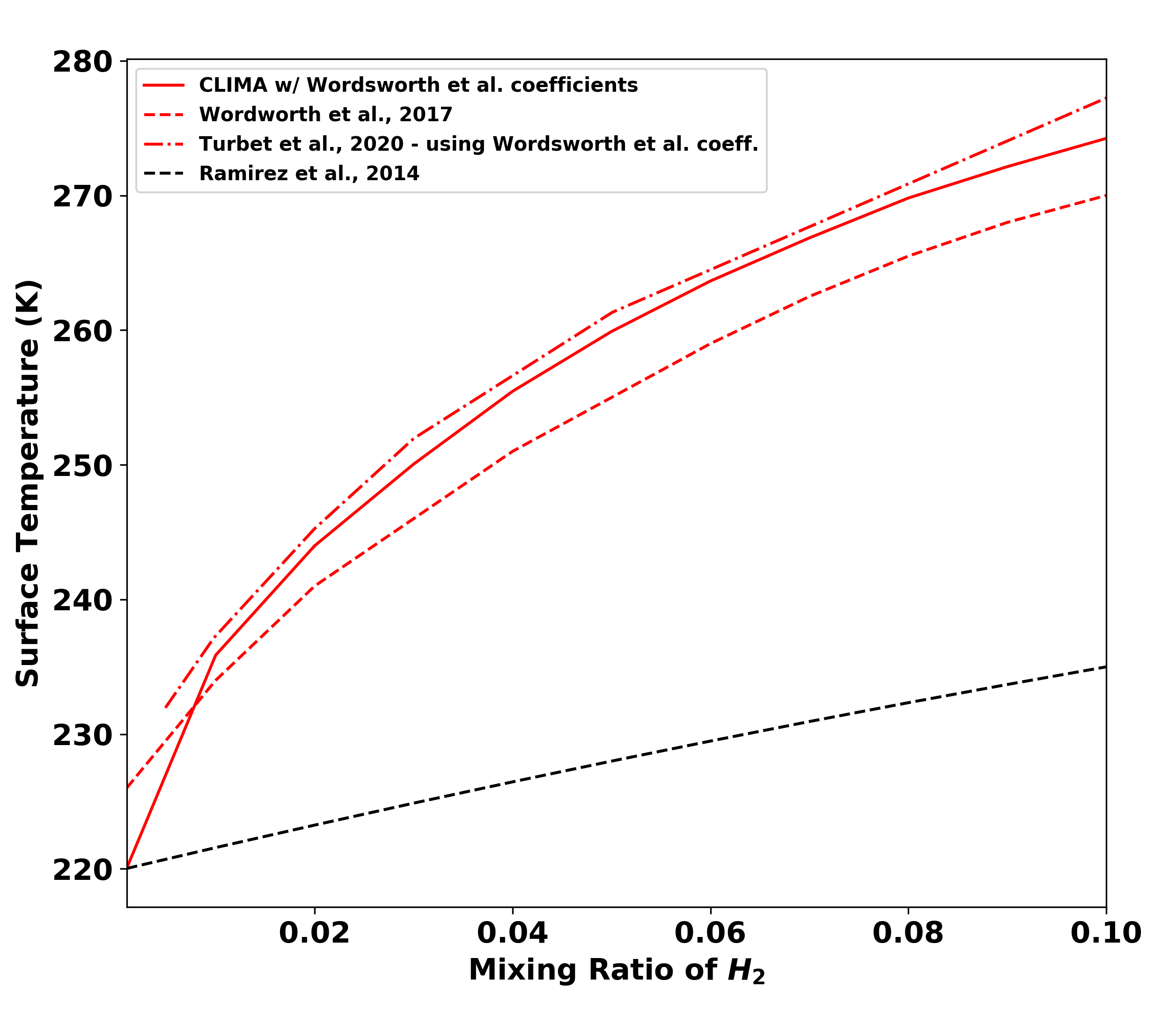}
 \end{center}
 \caption{Martian global mean surface temperature as a function of atmospheric $H_{2}$ mixing ratio, as calculated by our 1-D climate model. A 1-bar $CO_{2}$-$H_{2}O$ background atmosphere was assumed. Solar luminosity was 75 percent of present, appropriate for 3.8 Ga. The solid red curve was calculated using the CIA coefficients from \citet{wordsworth2017transient} The dashed red curve shows the results of \citet{wordsworth2017transient}, while the dashed-dot red curve corresponds to the results of \citet{turbet2019measurements} when using the CIA coefficients from \citet{wordsworth2017transient} The dashed black line shows temperatures calculated using the $N_{2}-H_{2}$ CIA coefficients from \citet{ramirez2014warming}}
 \label{fig:h2vstemp}
\end{figure}

\begin{figure*}[htbp] 
\begin{center}
 \includegraphics[width=6in]{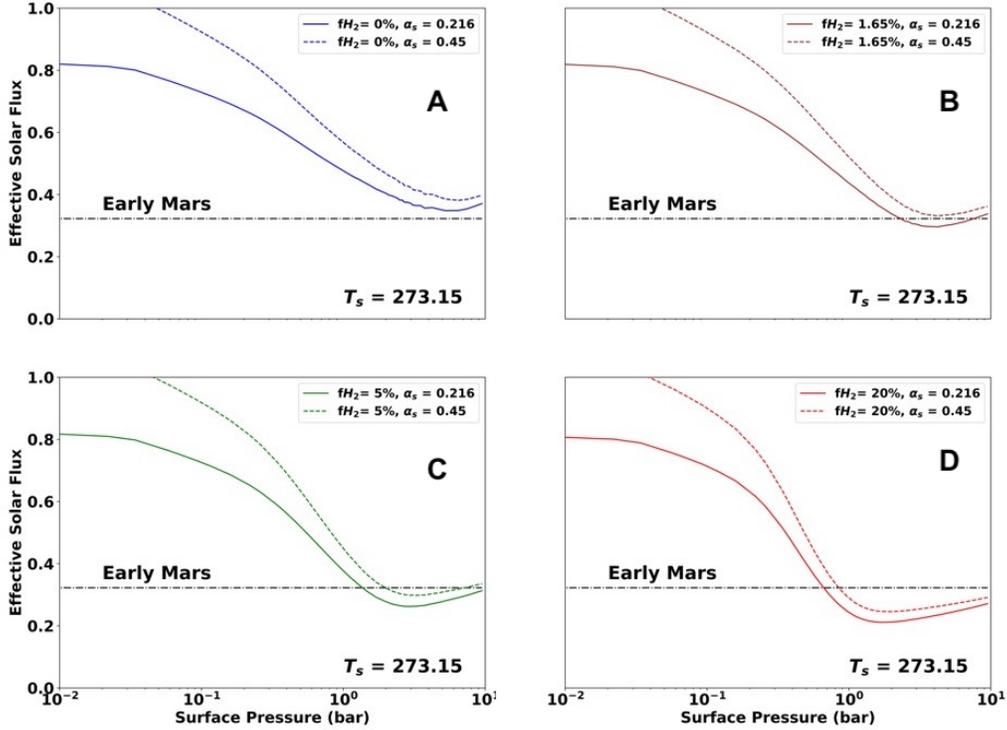}
 \end{center}
 \caption{Effective solar flux required to maintain a global mean surface temperature of 273.15 K for different surface pressures and atmospheric $H_{2}$ mixing ratios, as determined by our 1-D climate model. The dashed black line represents the effective solar flux (relative to modern Mars) at Mars' orbit at 3.8 Ga. We use two different surface albedos to simulate an ice-free ($\alpha_{s}$ = 0.216) and ice-covered ($\alpha_{s}$ = 0.45) surface. Figure A is for pure $CO_{2}$-$H_{2}O$ atmosphere, Figure B is for a $CO_{2}$ dominated atmosphere with $fH_{2}$ = 1.65\%, Figure C is for $fH_{2}$ = 5\%, and Figure D is for $fH_{2}$ = 20\%. If the mixing ratio line (colored) is able to cross below the solar flux line, the solution is physical.}
 \label{fig:4panel}
\end{figure*}

	We first used our 1-D model to model the warming produced by the new set of CIA coefficients. We did this by performing forward, time-stepping calculations with CLIMA for a 1-bar early martian atmosphere (see Figure~\ref{fig:h2vstemp}). The assumed surface albedo was 0.216, following \citet{ramirez2014warming}, this albedo is used to tune the model to modern Mars conditions. Surface temperatures, $T_{S}$, calculated using the CIA coefficients of \citet{wordsworth2017transient} are $\sim$30 K degrees warmer than those found by \citet{ramirez2014warming} using the old, $N_{2}-H_{2}$ coefficients. Recent work by \citet{turbet2019measurements} which includes the temperature dependence for $CO_{2}$-$H_{2}$ absorption shows that that their coefficients result in a colder surface temperature than when using the CIA coefficients from \citet{wordsworth2017transient}. Our original implementation of the CIA coefficients from \citet{turbet2019far} was flawed, so we will leave implementing their CIA coefficients to future work, but acknowledge that their coefficients would raise the amount of $H_{2}$ needed to bring $T_{S}$ above freezing (compared to our results). Our surface temperatures are $\sim$4 K higher than those calculated by \citet{wordsworth2017transient} for the same CIA coefficients, but almost equivalent to the surface temperature calculated by \citet{turbet2019measurements}. This is in part due to us assuming a fully saturated troposphere, while \citet{wordsworth2017transient} assumed a constant relative humidity of 0.8. We did not explore very low surface relative humidities as they are precluded by considerations of surface energy balance \citep{kasting2014remote}. This input discrepancy accounts for $\sim$2 K of the modeled difference. In our model, using the Wordsworth et al. CIA coefficients, the minimum $H_{2}$ mixing ratio required to bring $T_{S}$ above freezing at this $CO_{2}$ partial pressure is $\sim$0.09. 
  \begin{figure*}[htbp] 
\begin{center}
 \includegraphics[width=4.5in]{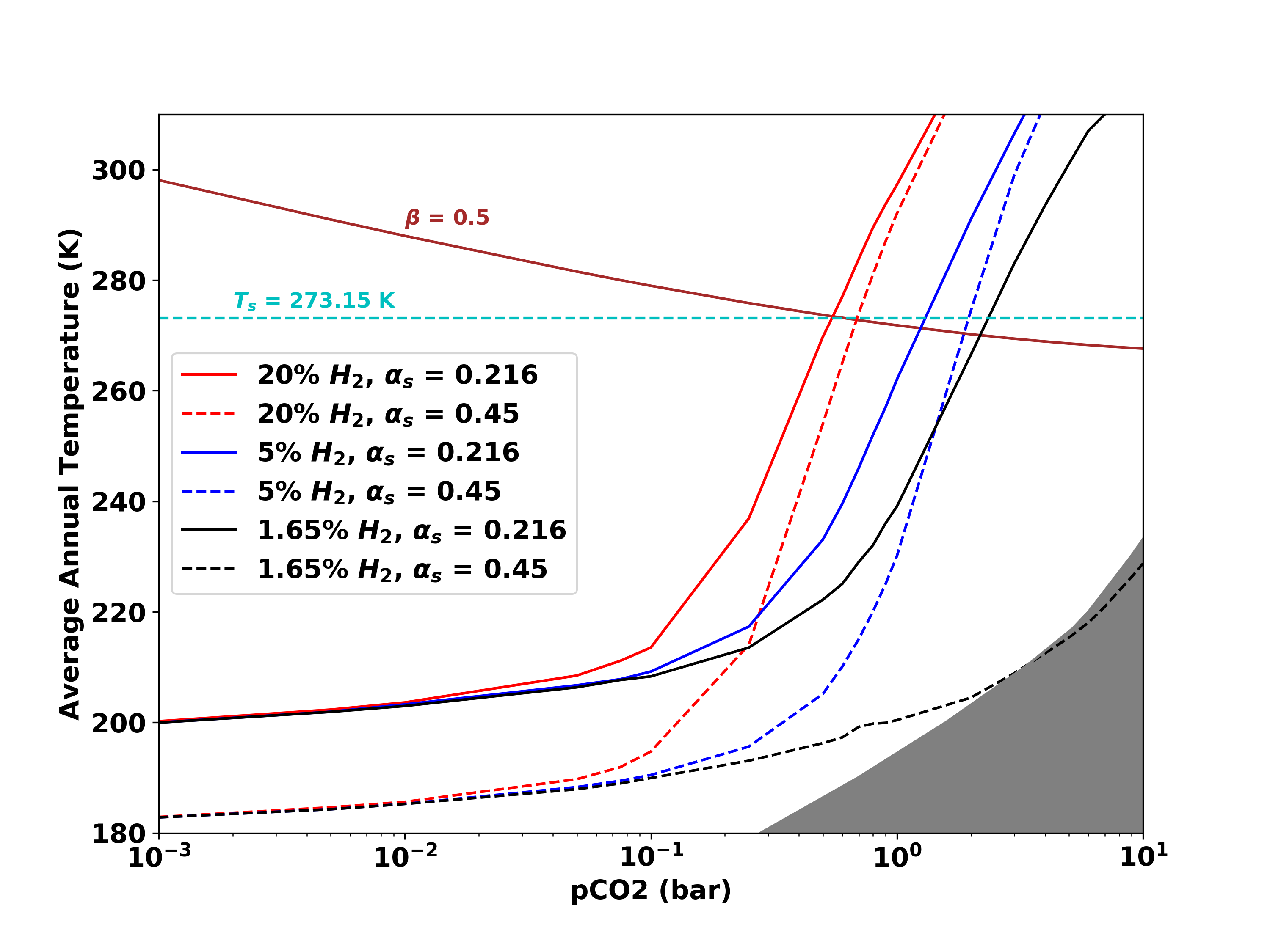}
 \end{center}
 \caption{Diagram illustrating when climate limit cycling should occur. The red, blue, and black greenhouse curves show surface temperature, $T_{S}$, as a function of $pCO_{2}$ for different atmospheric compositions (shown by color) and surface albedos (solid vs. dashed line).  The brown curve shows the temperature at which the weathering rate balances the present terrestrial $CO_{2}$ outgassing rate, for a given $pCO_{2}$ assuming $\beta$ = 0.5. If the greenhouse curves intersect the weathering curve below the freezing point, limit cycles will occur.  If the curves intersect above the freezing point, stable warm solutions should exist.  If the greenhouse curves never cross the freezing point, then the planet is permanently glaciated. The grey region in the plot is the pressure and temperature at which $CO_{2}$ begins condensing out of the atmosphere.}
 \label{fig:limitcycle}
\end{figure*}	
	Next, we ran CLIMA in inverse mode to solve for the effective solar flux, $S_{eff}$, as a function of surface pressure, $P_{S}$ (see Figure~\ref{fig:4panel}). For inverse climate modeling, we fix the atmospheric $P-T$ profile and then calculate the solar flux needed to sustain it, following \citet{kasting1991co2}. In this case, because we want to know how much $CO_{2}$ and $H_{2}$ are needed to stay above freezing, we fixed $T_{S}$ at 273.15 K.  Fig. 2 shows several things: First, it demonstrates that $CO_{2}$ alone cannot warm early Mars, as the pure $CO_{2}$ curve (blue) never intersects the black dash-dot line, which represents the solar flux for early Mars. Second, for most $H_{2}$ mixing ratios, the maximum warming occurs between $\sim$1 and 4 bars of $CO_{2}$. By contrast, for the pure $CO_{2}$ atmosphere, the maximum warming occurs at 6-7 bars. In our model we do appear to get slightly lower values for $S_{eff}$ than expected at high pressures (${S_{eff}}_{min}$= 0.83 rather ${S_{eff}}_{min}$= 0.86) \citep{kasting1991co2}. Though all the results in this paper are at pressures lower than at where this margin of disagreement occurs (less than 6 bars). The source of this disagreement is likely due to how we parameterize the absorption in the far wings of the $CO_{2}$ and $H_{2}O$ bands \citet{kopparapu2013habitable}. For $CO_{2}$ far-wing absorption, we use the "sub-Lorenzian" parameterization following \citet{perrin1989temperature}. And for $H_{2}O$ far-wing absorption, a "super-Lorenzian" parameterization is used from \citet{paynter2011assessment}. Further verification of these parameterizations will be left as future work. 
	
	We ran reverse calculations using a surface albedo of 0.216 to simulate an ice-free Mars, as well as for a surface albedo of 0.45 to simulate a glaciated Mars, following \citet{batalha2016climate}. As seen from the figure, solutions that have less than $\sim$5\% $H_{2}$ never dip below the black line when the planet is frozen. While it appears as though a 2\% $H_{2}$ case from Fig. 2 might be able to escape snowball, the continental configuration in the EBM affects surface albedo (see discussion Section~\ref{sec:discussion1}), so that model requires at least $\sim$6-8\% $H_{2}$ to escape snowball. An $H_{2}$ mixing ratio of 1.65\% (brown curves) corresponds to an $H_{2}$ outgassing rate of 64 Tmol/yr, which was the value favored by \citet{ramirez2014warming} based on analogy to Earth. Such an atmosphere would not be able to recover from snowball events and would thus fail to limit cycle.  

	We then used CLIMA to recreate Fig. 1 from \citet{batalha2016climate} using the new CIA coefficients from \citet{wordsworth2017transient}. The results are shown in Figure 4. Climate limit cycles should occur when the brown weathering rate curve (from eq. 3) intersects the greenhouse effect curves (red, blue and black curves) below the freezing point of water. We find that for an $H_{2}$ mixing ratio of 20\% (used in \citet{batalha2016climate}), we get permanently warm solutions. Our lower limit case of 5\% $H_{2}$ results in limit cycling. Solutions for mixing ratios between these two values fall right on the boundary between stable-warm and limit cycling solutions, so we turn to the EBM to explore their behavior. Calculated surface temperatures for the 1.65 percent $H_{2}$ case fall below the $CO_{2}$ condensation curve (which defines the boundary of the grey region in Figure 4) when the planet is fully glaciated, reinforcing the conclusion from Fig. 2 that such a mixing ratio is too low to produce even transiently warm climates. That said, calculations with HEx show that some latitudinal bands can remain above freezing even when the global average temperature is below zero. 

\subsection{HEx}
\label{sec:results2}
  \begin{figure}
\begin{center}
 \includegraphics[width=3.5in]{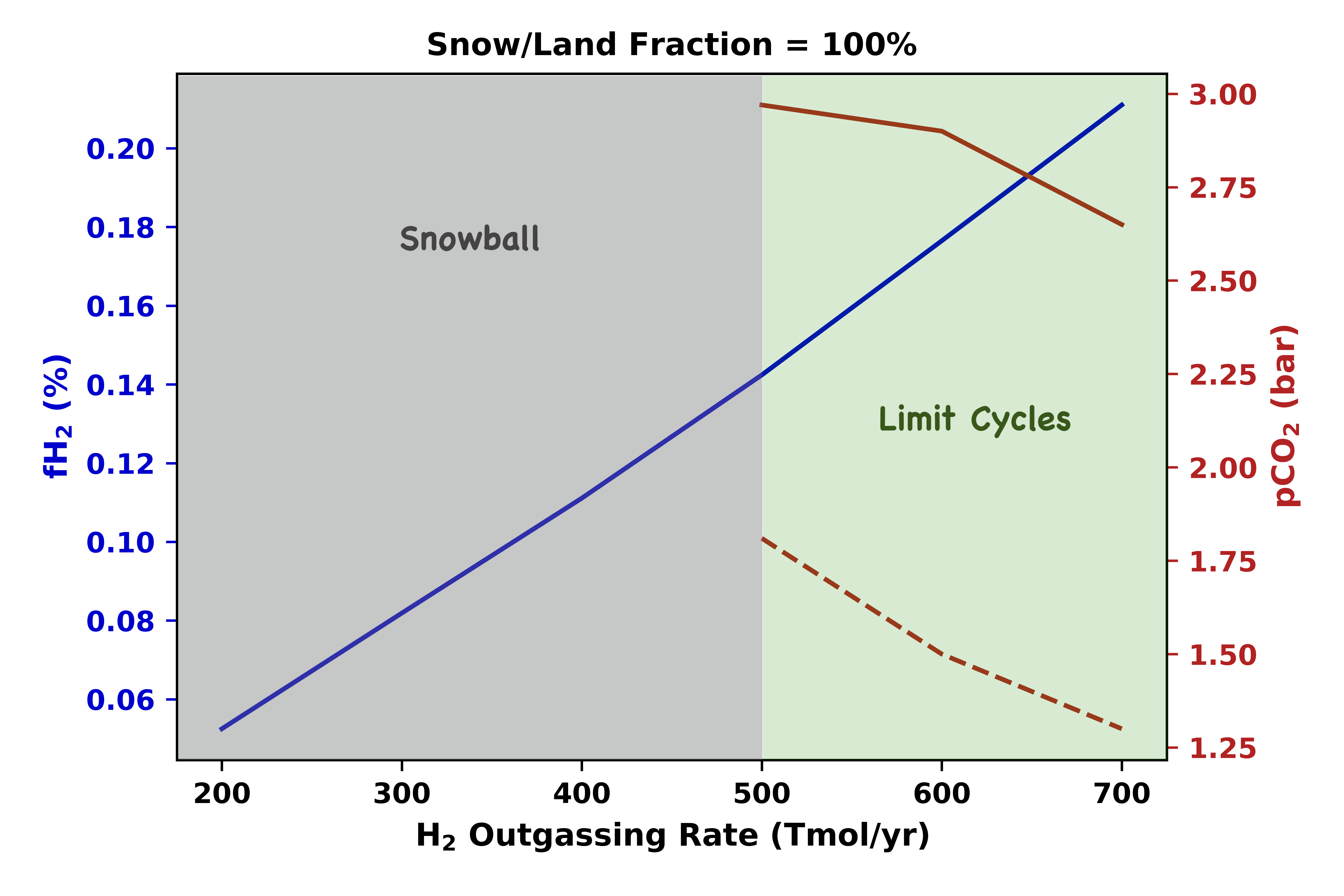}
 \end{center}
   \caption{This is for a $CO_{2}$ outgassing rate of 2.1 Tmol/yr. With the maximum snow/land coverage - a mixing ratio of $\sim$14\% is needed to escape snowball. All three figures show the resulting climate state as a function of $H_{2}$ outgassing rate and snow/land coverage. The blue curve shows the resulting mixing ratio of $H_{2}$ for each outgassing rate, while the red curve shows the resulting partial pressure of $CO_{2}$ to which the model equilibrates. The solid red curve is the maximum partial pressure of $CO_{2}$ and the dashed red curve is the minimum partial pressure of $CO_{2}$ that allows the planet to undergo limit cycle behavior.}
 \label{fig:ebm1}
\end{figure}

\begin{figure}
\begin{center}
 \includegraphics[width=3.5in]{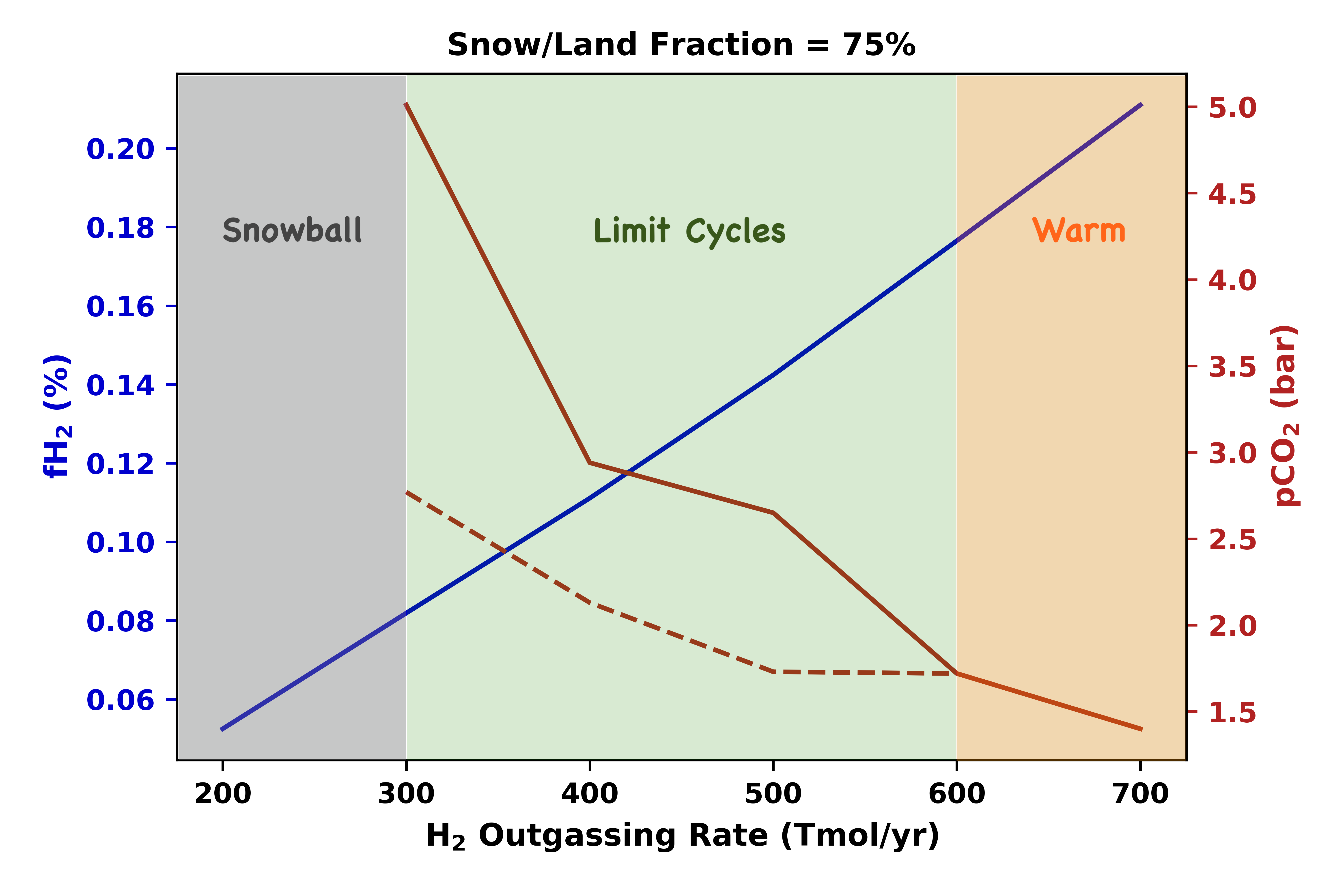}
 \end{center}
    \caption{This is for a $CO_{2}$ outgassing rate of 2.1 Tmol/yr. With a lower snow/land coverage of 75\%, a lower mixing ratio of $fH_{2}$ is required to escape snowball - approximately $\sim$8\%. The regime where limit cycles occur is smaller than the case where the ice-albedo feedback is stronger (See Figure~\ref{fig:ebm1}) - and stable warm states exist within the range of outgassing rates we explored.}
 \label{fig:ebm2}
\end{figure}

\begin{figure}
\begin{center}
 \includegraphics[width=3.5in]{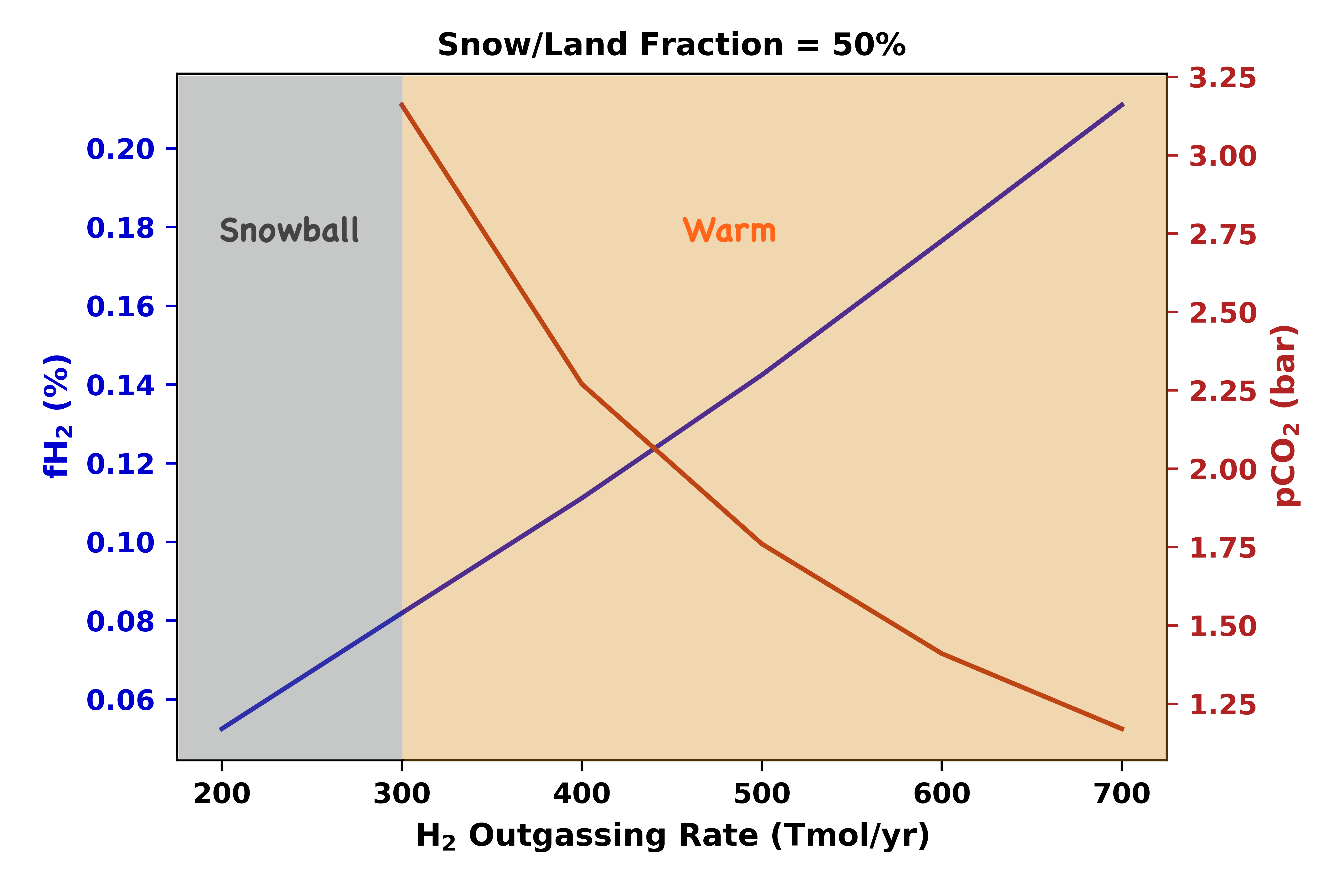}
 \end{center}
  \caption{This is for a $CO_{2}$ outgassing rate of 2.1 Tmol/yr. With only 50\% of frozen land covered in ice/snow - the planet loses the snowball bifurcation - resulting in either stable warm or permanent snowball states.}
 \label{fig:ebm3}
\end{figure}

	Using HEx, we varied the $H_{2}$ and $CO_{2}$ outgassing rates to explore the parameter space where limit cycling would be expected to occur. All runs started from a snowball state, as the ability to escape snowball is necessary for the climate to limit cycle. Figure 5-7 show at what $H_{2}$ outgassing rates we expect snowball, limit cycling, and permanently warm states - as well as the corresponding quantities of $H_{2}$ and $CO_{2}$ for that climate state. As these results are highly sensitive to any potential error in our PALB (planetary albedo) polynomial fits, so we also varied the snow/land fraction for our glaciated planet. This not only allowed us to see how potential errors in PALB may affect limit cycling behavior, but also the effect limited surface water has on limit cycling in general. As we moved from higher to lower snow/land fractions (see Figure~\ref{fig:ebm1},\ref{fig:ebm2},\ref{fig:ebm3}), there are two key takeaways. First, the weaker ice-albedo feedback allows us to deglaciate with lower $H_{2}$ outgassing rates - as was pointed out by \citet{ramirez2017warmer}. Second, the weaker ice-albedo feedback eventually removes the snowball bifurcation, eliminating limit cycling altogether. As a caveat, it should be noted that we assume the ratio of snow to land to be identical between every latitudinal band. A more sophisticated treatment of the hydrologic cycle may yield higher snow/land ratio's in the highlands \citep{wordsworth2013global} or near the poles compared to the rest of the planet - but we leave this as future work. Further study would include more detailed modeling of how water availability would impact the ice-albedo feedback and subsequently limit-cycling behavior on an early Mars. 
	
	Figure 8 shows the fraction of time any portion of the surface remains above the freezing point of water as a function of $H_{2}$ and $CO_{2}$ outgassing rates. For the parameter range we explored, $CO_{2}$ outgassing was a control on frequency of limit cycles than existence of limit cycling. Higher $CO_{2}$ outgassing rates allowed the planet to escape snowball faster, as well as increasing the duration of the transient warm period. The snow/land fraction and the $H_{2}$ outgassing rate are the controls on whether limit cycles occur for the parameter space we explored, see Figure~\ref{fig:ebm1},\ref{fig:ebm2},\ref{fig:ebm3} for the cutoffs between permanent snowball, limit cycling, and permanent warm climate states. 

  \begin{figure}
\begin{center}
 \includegraphics[width=3.5in]{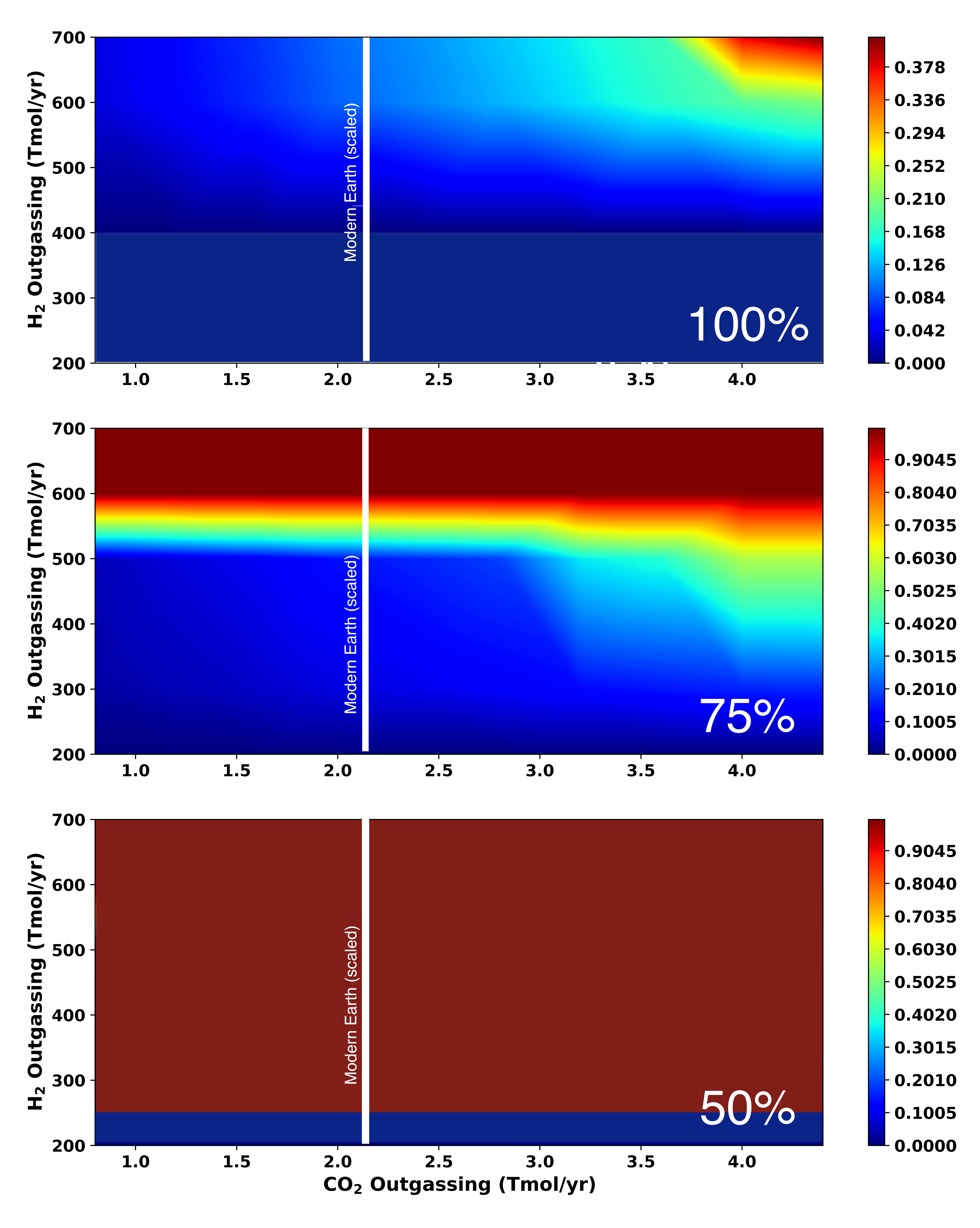}
 \end{center}
  \caption{These three panels show the fraction of time the planet remains above the freezing point of water as a function of $CO_{2}$ and $H_{2}$ outgassing rates and snow/land coverage (shown in white text in bottom right). As the ice-albedo feedback mechanism is weakened by lower snow coverage, lower $H_{2}$ values can be used to escape snowball, but the limit cycling regime becomes smaller. For reference - terrestrial outgassing rates of $CO_{2}$ and $H_{2}$ scaled for Mars' surface area and mantle fugacity are $CO_{2}= 2.1\frac{Tmol}{yr}$ (designated by the vertical white line) and $H_{2}= 64\frac{Tmol}{yr}$.}
 \label{fig:ebm4}
\end{figure}
	
\begin{figure*}[htbp] 
\begin{center}
 \includegraphics[width=6in]{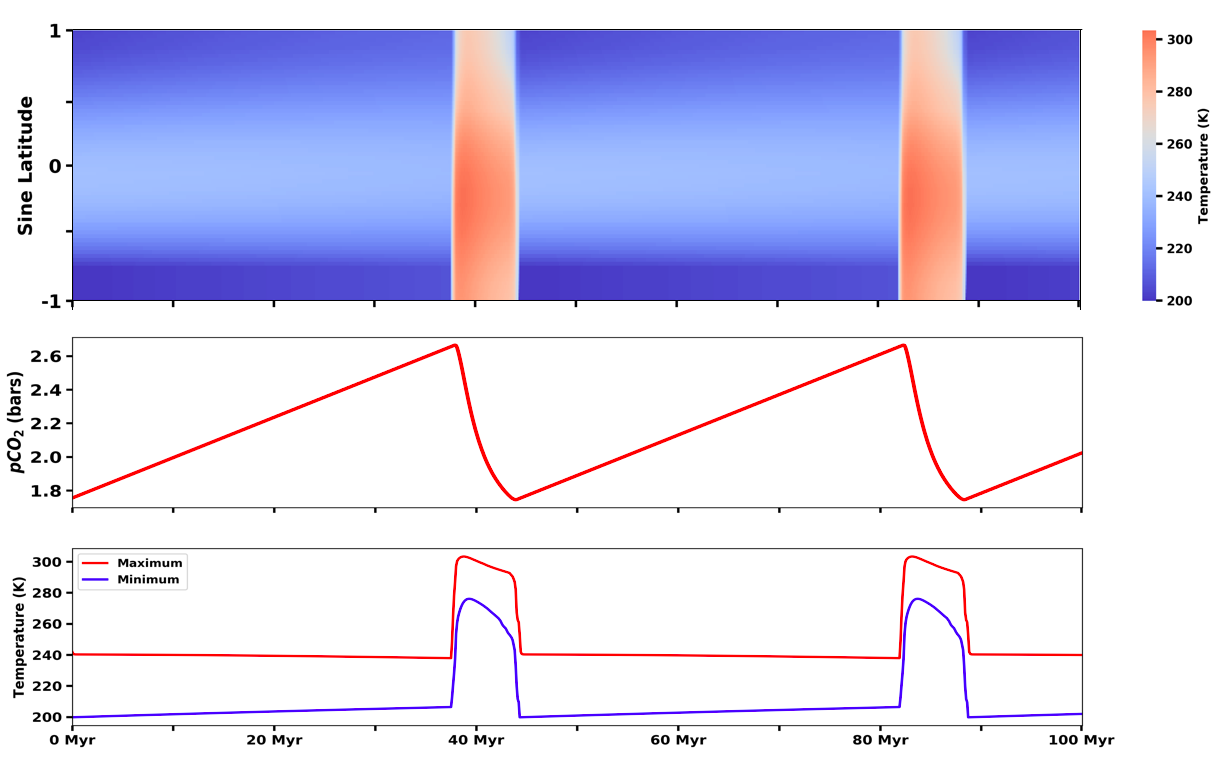}
 \end{center}
  \caption{This plot shows the average annual surface temperature as a function of latitude over time for the outgassing rates of $CO_{2} = 2.0\frac{Tmol}{yr}$ and $H_{2} = 500\frac{Tmol}{yr}$ with 75\% snow/land fraction during glaciated states.}
 \label{fig:ebm5}
\end{figure*}

	To illustrate the limit cycling behavior more clearly, we plotted the average annual temperature by latitude over time for the 500 Tmol/yr outgasing case of an $H_{2}$ with 75\% snow/land coverage (Figure~\ref{fig:ebm5}). This particular case was run for 100 Myr with two transient warm periods (each approximately 6 Myr in duration) separated by prolonged snowball periods. Though the cases are not plotted here, as one lowers the $H_{2}$ outgassing rate, warm periods can reach lengths of $\sim$100 Myr while also lengthening the duration of the snowball state. Increases to the $CO_{2}$ outgassing rate work to shorten snowball periods, but have only a small impact on the length of warm climate states. 
		
\section{Discussion}
\label{sec:discussion}

\subsection{The Ice-Albedo Problem}
\label{sec:discussion1}

	While a strong ice-albedo feedback is the driving reason for an instability existing between stable climatic states \citep{budyko1969effect}, it can also cause difficulties with warming a planet warm in the first place \citep{ramirez2017warmer,ramirez2018geological}. \citet{ramirez2017warmer} showed that a frozen, early Mars-like planet with $fH_{2}$ = 10\% and a  surface albedo of 0.55, or $fH_{2}$ = 20\% and a surface albedo of 0.65, would be incapable of escaping the snowball. At a slightly lower surface albedo of 0.45, our model reaches the freezing point of water at relatively low values of $fH_{2}$ (Fig. 2). So, a key question is: What is the surface albedo of a fully glaciated early Mars-like planet? In our previous work, \citet{batalha2016climate} assumed a snow albedo of 0.45 with 100\% surface coverage when the planet is glaciated. We mimicked this by assuming a snow albedo of 0.45 and varied the snow surface coverage of frozen latitudes. As shown in Section 3.2, the assumptions one makes for ice-albedo feedback greatly impact the ability of a planet to limit cycle. 
	
	\citet{ramirez2017warmer} argued that limit cycling cannot occur at modest $H_{2}$ levels for substantial clean ice/snow coverage. This is a valid point. However, assuming clean snow coverage over large swaths of the Martian surface may be unrealistic.  Unlike Snowball Earth, a glaciated early Mars was likely water-limited \citep{wordsworth2013global}, allowing for only partial ice coverage.  Partial ice coverage would not only lower the overall planetary albedo, but it could shut off precipitation to exposed continental areas, resulting in local climates similar to that of the McMurdo dry valleys on Earth \citep{head2014climate}. Three-dimensional and one-dimensional modeling have shown that for Earth-like cases, exposure of these dry continental areas introduces dust particles into the atmosphere, which then accumulate elsewhere on the planet \citep{abbot2010mudball,hoffman2017snowball}.  Atmospheric dust does two things that may help warm the climate: i) It contributes to the greenhouse effect when the planetary surface albedo is high \citep{abbot2010dust}. And, ii) it lowers the albedo of glaciated regions by creating 'dirty' snow \citep{conway1996albedo}.
	
	We conclude that the relatively high albedo of ice and snow does indeed raise issues in keeping early Mars warm; however, these issues do not seem to be insurmountable. Perhaps more importantly, just pointing out that such problems exist is not sufficient to rule out limit cycling behavior, as the warm conditions favored by \citet{ramirez2017warmer} and \citet{ramirez2018geological} are only stable if $CO_{2}$/$H_{2}$ outgassing rates were sufficiently high and/or the ice-albedo feedback was sufficiently weak (See Figure~\ref{fig:ebm4}). 
	
\subsection{Stagnant Lid Tectonics and Supply-Limited Weathering}
\label{sec:discussion2}

	Our weathering parameterization is based on one developed for Earth, and hence it implicitly assumes that plate tectonics was operating on early Mars. This may or may not have been true \citep{kerr1999signs}. Alternatively, Mars may always have been in a stagnant lid tectonic regime, similar to that which exists there today \citep{breuer2003early}. A potential critique of the limit-cycling hypothesis faced by stagnant lid planets is availability of $CO_{2}$ for outgassing over extended periods of time.  Unlike planets with plate tectonics to deliver $CO_{2}$ back into the mantle, stagnant lid planets will eventually deplete the mantle of $CO_{2}$ and sequester it in the lithosphere \citep{foley2018carbon,foley2019habitability}. However, point volcanism and metamorphic outgassing may continue to cycle $CO_{2}$ back into the atmosphere even after the mantle is depleted of $CO_{2}$ \citep{pollack1987case,kerrick2001metamorphic}. Consider a conservative case, in which $CO_{2}$ outgassing is purely from the mantle and in which carbonates, once formed, never release their $CO_{2}$ back into the atmosphere. If the total initial carbon reservoir was $200 \times 10^{18}$ kg \citep{grady2006carbon}, $CO_{2}$ could have been outgassed at a constant rate of 2.1 Tmol/yr for $\sim$7.9 Gyr, if volcanism remained active the entire time. This suggests that limit cycling behavior  would have been constrained by the lifetime of volcanism on early Mars, rather than by carbon availability. Any recycling of carbonates, for example by point volcanism \citep{pollack1987case}, would only strengthen this conclusion. Future work to try to quantify this behavior might include modeling the thermal evolution of the Martian mantle and coupling the variable outgassing flux of $CO_{2}$ to our climate model. This could qualitatively change our results, as a high $CO_{2}$ outgassing rate early in Mars' history could initially have resulted in stable warm climates. These would be followed by a period of limit cycling as the martian mantle cooled and volcanism decreased, until Mars reached its current climate state. 
	
	The weathering rate might also have been quite different on an early Mars-like planet in a stagnant lid regime, as such a planet would lack the constant resupply of weatherable material to the surface that we find on modern Earth. This 'supply-limit' to the weathering rate can allow massive $CO_{2}$ greenhouses to form - affecting a planet's ability to regulate its climate by disrupting the carbonate-silicate cycle \citep{mills2011timing,foley2015role}. We estimate the weathering supply limit for Mars using the following expression from Foley (2015; 2019):
\\
\begin{equation}
F_{sl} = \epsilon f_{m} \chi \rho_{l}
\end{equation}
\\
Here, $\epsilon f_{m}$ is the eruption rate [$\frac{m^{3}}{yr}$], $\chi$ is the weathering demand for complete carbonation of basalt [$\frac{mol}{kg}$], and $\rho_{l}$ is the density of our material [$\frac{kg}{m^{3}}$]. \citet{o2007melt} estimated volumetric eruption rates throughout Mars' thermal history and found a rate of 0.17 $\frac{km^{3}}{yr}$ for the early Hesperian (the furthest back they extrapolated with their model). This should be treated as a lower limit, as it was calculated for the early Hesperian, not the late Noachian. If we assume a weathering demand of 5.8 $\frac{mol}{kg}$ \citep{foley2019habitability} and a melt density of 3000 $\frac{kg}{m^{3}}$ \citep{o2007melt}, the stagnant lid supply-limited weathering rate is $\sim$$3 \times 10^{12}$ $\frac{mol}{yr}$. By comparison, the maximum and minimum weathering rates over the warm periods during our EBM calculations is between $\sim$$1.7 \times 10^{13}$ $\frac{mol}{yr}$ and $\sim$$2.2 \times 10^{13}$ $\frac{mol}{yr}$. This suggests that silicate weathering would have been supply-limited, given the eruption rates  from \citet{o2007melt}. But eruption rates greater than their early-Hesperian value could have supplied more fresh weatherable rock to take up the volcanic $CO_{2}$, and thus regulate early Mars' climate. Additionally, we are hitting the supply-limited weathering rate post glaciation which is a transient phase that results in a prolonged-hot climate. But as surface temperature decrease the weathering rate will fall back below the supply-limit. Future work would involve modeling this unique weathering limit and how it would impact the climate of early Mars and the limit cycling hypothesis. 

\subsection{Obliquity}
\label{sec:discussion3}

	Recent constraints on Mars' obliquity history from analysis of cratering patterns in the martian geomorphological record puts limits at approximately 10$^{\circ}$ and 40$^{\circ}$ obliquity \citep{holo2018mars}. Though not treated as one of our primary free parameters (unlike volcanic outgassing rates), we tested this range of obliquity solutions in our model. 
	
	For stable-warm solutions, we found that the lower limit on $fH_{2}$ was unaffected by changes in obliquity and only changed the position of the planet's ice-line. Current model limitations keep us from studying high obliquity cases when the planet is completely glaciated - barring us from exploring obliquities effect on limit cycles. Seasonal condensation and sublimation of $CO_{2}$ are not currently incorporated into our $CO_{2}$ cycle, and will be added as future work. Currently, our model keeps $CO_{2}$ from condensing at the poles by raising the temperature to the $CO_{2}$ saturation temperature for the given pressure - similar to that done by \citet{batalha2016climate}. A more accurate treatment of the carbon cycle would be warranted for exploring the extreme obliquity cases

\subsection{Comparison with other geologic constraints}
\label{sec:discussion4}

$i) \ Surface \ pressure$: \citet{kite2019geologic} cites an upper boundary on surface pressure of $\sim$1 bar \citep{kite2014low,williams2014production}. The upper bound is derived from the size distribution of primary impact craters. This places all our climatically acceptable results ($\sim$1.2 to $\sim$3 bars) above the upper limit derived from such data - though is an improvement from our previous results of $\sim$3 to $\sim$4 bars \citep{batalha2016climate}.

$ii) \ Timescales$:  Sediment transport needed to form the observed valley networks during the late Noachian requires a cumulative $\sim$$10^{6}$ years or more of "wet" climate \citep{kite2019geologic,hoke2011formation,orofino2018estimate}. Our limit cycling calculations produced warm periods that lasted on the order of $\sim$$10^{6}$ years each. So, the number of cycles would not be a limiting factor on cumulative wet years needed to form the valley networks, as even one cycle is sufficient to reach the minimum duration required by \citet{kite2019geologic}.

$iii) \ CO_{2}$ $and~Carbonates$:  A longstanding puzzle for warming early Mars' with a predominately $CO_{2}$ atmosphere has been the lack of carbonate deposits on the martian surface \citep{bibring2006global}. We propose that the high $CO_{2}$ partial pressure acts to acidify the rainwater, which in turn dissolves any surface carbonates present \citep{kasting2012find}. The carbonate would then be redeposited in the martian subsurface as the rainwater percolates downward and its acidity was buffered by the abundant basalt. This process would only be exacerbated by the presence of sulfates in the rainwater - further lowering the rainwater's $pH$. Evidence exists for highly acidic surface conditions \citep{ming2006geochemical,hurowitz2006situ}, though further modeling is needed to quantify this theory and its application to the carbonate $mystery$.

$iv) \ H_{2} \ outgassing \ rates$: Although the required $H_{2}$ mixing ratios and outgassing rates estimated here are a factor of two lower than those found by \citet{batalha2016climate}, they still require volcanic outgassing rates that are at least four times the outgassing rate on modern Earth (scaled for planet size and for mantle oxygen fugacity).  This is the same conclusion as that reached by \citet{ramirez2014warming} using a simpler, 1-D climate model. (Coincidentally, the enhanced warming from the new $H_{2}$-$CO_{2}$ CIA coefficients is almost exactly offset by the enhanced surface albedo caused by the existence of globally glaciated states in the EBM.) For those who think that such outgassing rates are too high for early Mars, we offer three possible ways by which their magnitudes might be reduced: 

	1) The first is that $H_{2}$ may have been escaping at less than the diffusion-limited rate. A new paper by \citet{zahnle2019strange} shows that hydrodynamic escape of $H_{2}$ from early Earth may have been several times slower than the diffusion limit, provided that the solar XUV flux was not too high. Whether this might also have been true for early Mars is not clear. The solar XUV flux at 3.8 Ga is estimated to be higher than today by a factor of $\sim$40, according to \citet{zahnle2019strange} eq. (2). The fact that Mars is less massive than Earth (and thus requires less energy for escape) is roughly compensated by its greater distance from the Sun. Bearing this in mind, their Fig.5 suggests that martian $H_{2}$ could have escaped at the diffusion limit. But this result is from a 1-D escape model. The actual escape rate could have been slower if geometry and magnetic fields played an important role \citep{stone2009anisotropic} and if $CO_{2}$ or other radiatively active constituents (e.g. $CO$, $H_{2}O$, $O$) acted as a coolant - scavenging XUV radiation and further slowing hydrodynamic escape.

	2) A second solution is that the estimated $H_{2}$ ougassing rate for early Mars may be overly conservative \citep{ramirez2014warming}. These authors were concerned with direct volcanic outgassing. Other mechanisms for $H_{2}$ production have been proposed with the most recent being magnetite authigenesis \citep{tosca2018magnetite}. In anoxic and high pH environments, basaltic minerals dissolved in water are kinetically driven to precipitate out magnetite and release $H_{2}$, which then exsolves from the solution. The authors claim that this production method can outpace the volcanic outgassing rate of $H_{2}$ and alone could account for a several percent mixing ratio of $H_{2}$. 
	
	3) A third option is that an additional greenhouse agent assisted in trapping OLR, perhaps $CH_{4}$. It should be noted that conventional absorption of IR radiation by methane (i.e., by permitted rovibrational transitions) fails to produce warming in a $CO_{2}$-rich atmosphere; instead it cools the planet via its anti-greenhouse effect \citep{ramirez2014warming,wordsworth2017transient}. Methane can, however, produce greenhouse warming by way of CIA with $CO_{2}$ \citep{wordsworth2017transient,turbet2019far}. Accumulating enough $CH_{4}$ to contribute substantially to warming via CIA may be difficult on early Mars, as abiotic serpentization has been experimentally shown to only convert 0.04\% of outgassed $H_{2}$ to $CH_{4}$, assuming 100\% serpentinite \citep{oze2005have,tarnas2019insufficient}. As $CO_{2}$-$CH_{4}$ CIA is considerably weaker than $CO_{2}$-$H_{2}$ CIA, warming the planet with methane would require either much higher mixing ratios than for $H_{2}$ or significantly higher surface pressures to achieve the same greenhouse forcing. Alternative mechanisms have been suggested to trap methane produced via serpentization within clathrates, releasing it in bursts - potentially triggered by atmospheric collapse and obliquity shifts \citep{chassefiere2016early,wordsworth2017transient,kite2017methane}. Methane in this capacity may have assisted $H_{2}$ as a warming agent, but its invocation as a major greenhouse gas is highly conditional and possibly unnecessary. 
	
\section{Conclusion}

Using the $CO_{2}$-$H_{2}$ CIA coefficients of \citet{wordsworth2017transient} in our EBM climate model, we showed that climate limit cycles to occur at relatively modest $CO_{2}$ and $H_{2}$ concentrations on early Mars \citep{kite2019geologic}. Our climate model is able to deglaciate the planet at 3.8 Ga at $fH_{2}$ values as low as 5 to 8\% for surface pressures between 1.2 and 3 bars. We also found these results to be highly sensitive to how the ice-albedo feedback is parameterized. This warrants future work to look into water availability and it's effect on the planetary albedo of a frozen Early Mars. This type of intermittent warming is broadly consistent with various interpretations of Mars' geology that suggest intermittent warm and cold climate conditions.

\section*{Code Availability}
Both of the codes (CLIMA and HEx) used in this paper are available upon request from the corresponding author. The OLR and PALB fits are available with the online version of the paper. 

\appendix
\section{Polynomial fits to OLR and Planetary Albedo}
\label{appendix}
We parameterized top of the atmosphere (TOA) albedo, $\alpha$, and the outgoing IR flux, $F_{OLR}$, as
polynomials with the following variables:
surface temperature $T(K)$ used as $t=Log10(T(K))$, $\phi = Log10(P_{s})$, where $P_{s}$ is the
surface pressure,  $fh_{2}$ is the volume mixingratio of H$_{2}$, $\mu = cos(z)$
where $z$ is the zenith angle, and $a_{s}$ is the surface albedo.
The parameterizations were derived by running the 1-D radiative convective (RC) model over a range of values of
the above parameters for each stellar type. 
The fits are valid in the range $10^{-3}< P_{s} < 10$ bar, $0 < fh_{2} < 0.2$, $0 < a_{s} < 1$, $0 < \mu < 1$ and
$210 K \le T \le 350 K$. For both the albedo and OLR, we split the fits as such for better accuracy: 210 K - 230K, 
230 K - 250 K, 250 K - 270 K, 270 K - 290 K, 290 K - 310 K, 310 K - 330K.

The TOA parameterizations indicate that TOA should increase with $Z$. In our 1-D RC model, the 
stratospheric temperature $T_{strat}$ is calculated as follows:
\begin{eqnarray} 
 T_{\mathrm{strat}} &=& \frac{1}{2^{1/4}}\left[ \frac{S}{4 \sigma} (1 - \alpha)\right]
\end{eqnarray} 
where S is incident solar flux, and $\sigma$ is the Stefan-Boltzmann constant. 
This equation,
however, is appropriate only for global average conditions (i.e., for $Z$ $= 60^{\circ}$) 
and can not be applied to calculate $T_{strat}$ and $\alpha$ for latitudes
having solar zenith angles different from this value.
Following Williams \& Kasting \citep{williams1997habitable}, we parameterized $T_{strat}$ as a function of $Z$.
\begin{eqnarray} 
T_{\mathrm{strat}}(Z) &=& T_{strat} (60^{\circ}) \left[ \frac{F_{s}Z)}{F_{s}(60^{\circ})}\right]^{1/4}
\label{tstrat}
\end{eqnarray} 
where $F_{s}$ is the absorbed fraction of incident solar
flux, which was calculated for a variety of zenith angles between $0^{\circ}$ and $90^{\circ}$ using the
radiative-convective model, and Tstrat($60^{\circ}$) was obtained using Eq.(\ref{tstrat}) above.

\subsection{Error Estimates in OLR}

Below we have quantified the percent error in our OLR fits for select temperature ranges (for reference).
  \begin{figure}[h!] 
\begin{center}
 \includegraphics[width=6in]{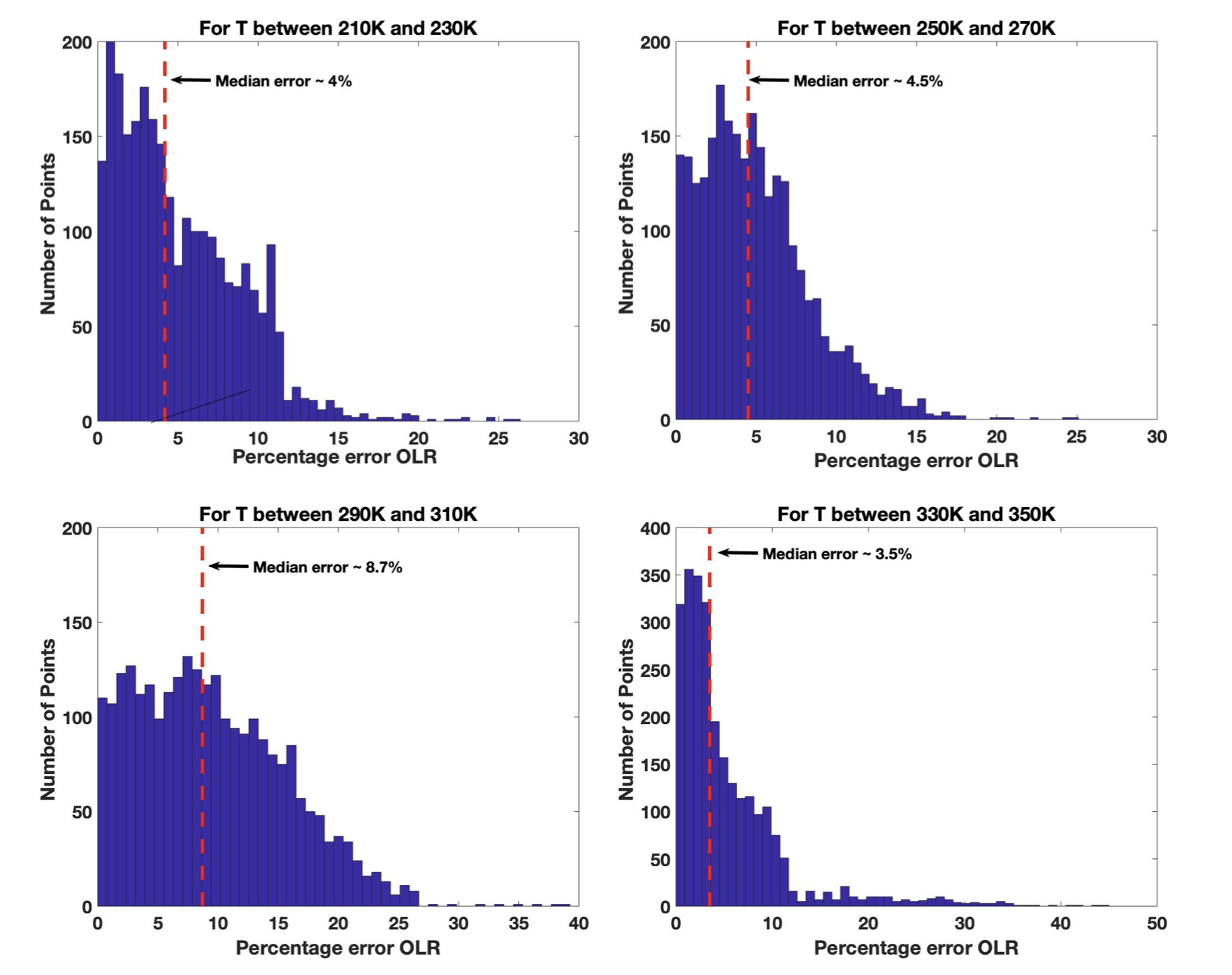}
 \end{center}
 \label{fig:error1}
\end{figure}

\newpage
\subsection{Error Estimates in PALB}

Below we have quantified the percent error in our PALB fits for select temperature ranges (for reference).
  \begin{figure*}[h!] 
\begin{center}
 \includegraphics[width=6in]{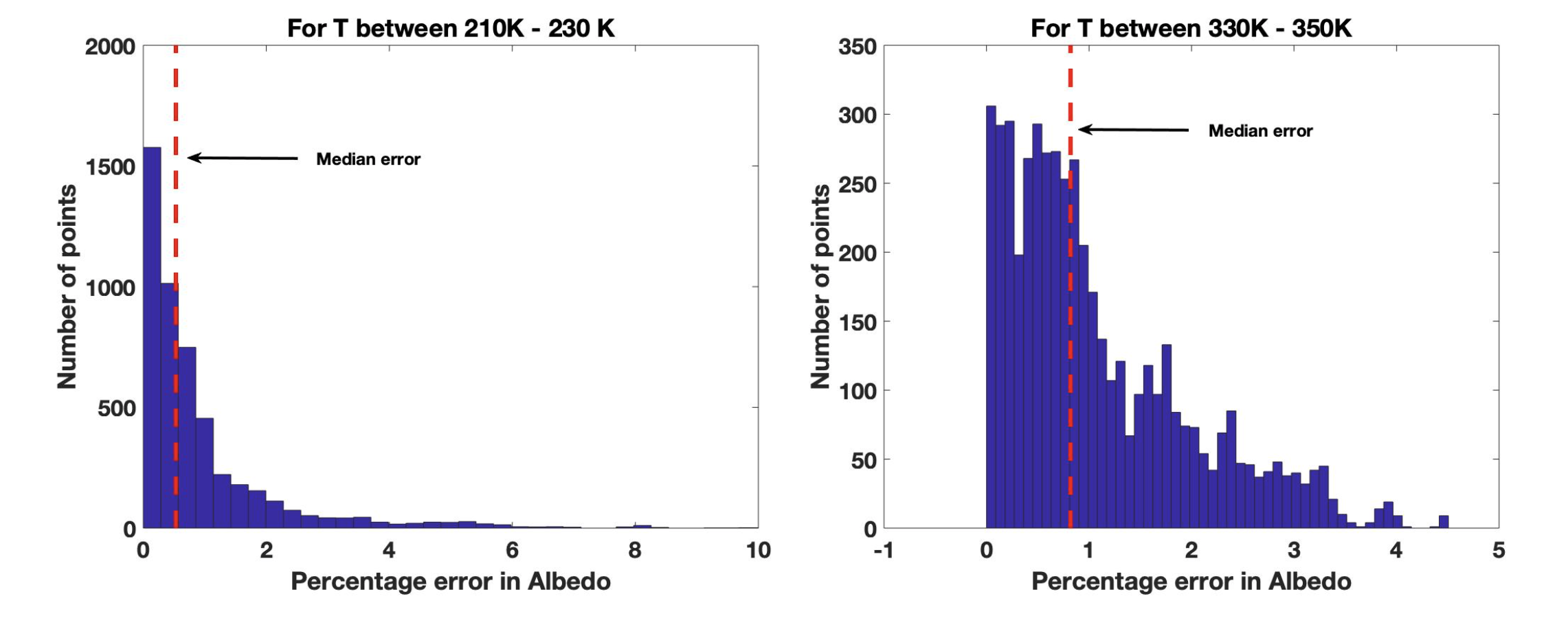}
 \end{center}
 \label{fig:error5}
\end{figure*}

\section*{Acknowledgements}

We would like to thank both of our anonymous reviewers for their invaluable feedback. We would also like to thank Edwin Kite for insightful discussion on martian paleoclimate.


\bibliography{bib}

\end{document}